\documentstyle[aps,floats,prd,psfig]{revtex} 
     
\newcommand{\ApJ}{Astrophys. J.}

\newcommand{\da}{d_A}

\newcommand{\lin}{{}}
\newcommand{\bn}{\hat{\bf n}}    
\newcommand{\bx}{{\bf x}}

\newcommand{\veck}{\hat{\bf k}}

\newlength{\tskip}
\newlength{\colwidth}

\newcommand{\beq}{\begin{equation}}
\newcommand{\eeq}{\end{equation}}
\newcommand{\beqa}{\begin{eqnarray}}
\newcommand{\eeqa}{\end{eqnarray}}

\long\def\comment#1{}
\newcommand{\wj}{\left(
                          \begin{array}{ccc}
                          l_1  &  l_2  & l_3 \\
                            0  &  0    &  0
                          \end{array}
                          \right)}
\newcommand{\wjm}{\left(
                           \begin{array}{ccc}
         l_1 & l_2  & l_3  \\
         m_1 & m_2  & m_3
                           \end{array}
                   \right)}

\newcommand{\deld}{\delta^{\rm D}}

\newcommand{\bk}{{\bf k}}
\newcommand{\rad}{r}    

\newcommand{\Ylm}[1]{Y_{l_#1}^{m_#1}}

\newcommand{\Ytwotwom}[1]{_{\pm 2} Y_{2}^{m_#1}}
\newcommand{\Ylmn}{Y_{l}^{m}}
\newcommand{\Ytwomn}{Y_{2}^{m}}
\newcommand{\Ytwolmn}{_{\pm 2} Y_{l}^{m}}
\newcommand{\Ytwotwomn}{_{\pm 2} Y_{2}^{m}}

\newcommand{\Dk}{\frac{d^3{\bf k}}{\left( 2\pi \right) ^3}}

\begin{document}   

\twocolumn[\hsize\textwidth\columnwidth\hsize\csname @twocolumnfalse\endcsname
\title{Cross-Correlation Studies with CMB Polarization Maps}
\author{Asantha Cooray}
\address{California Institute of Technology, Mail Code
130-33, Pasadena, CA 91125\\
E-mail: asante@tapir.caltech.edu}

\maketitle
\begin{abstract}
The free-electron population during the reionized epoch rescatters CMB temperature quadrupole and generates
a now well-known polarization signal at large angular scales. While this contribution has been detected in
the temperature-polarization cross power spectrum measured with WMAP data, due to the large cosmic variance 
associated with anisotropy measurements at tens of degree angular scales
only limited information related to reionization, such as the optical depth to electron scattering, can be extracted.
The inhomogeneities in the free-electron population lead to an additional secondary polarization anisotropy contribution 
at arcminute scales. While the fluctuation amplitude, relative to dominant primordial fluctuations, is small, 
we suggest that a cross-correlation between arcminute scale CMB polarization data and a tracer field of the high redshift universe, 
such as through fluctuations captured by the 21 cm neutral Hydrogen background or those in the
infrared background related to first proto-galaxies, may allow one to study additional details related to reionization.
For this purpose, we discuss an optimized higher order
correlation measurement, in the form of a three-point function, including 
information from large angular scale CMB temperature anisotropies in addition to 
arcminute scale polarization signal related to inhomogeneous reionization.  
We suggest that the proposed bispectrum can be measured with a substantial signal-to-noise ratio
and does not require  all-sky maps of CMB polarization or that of the tracer field.
A measurement such as the one proposed may allow one to establish the epoch when CMB polarization related to reionization is generated and to
address if the universe was reionized once or twice.
\end{abstract}

\hfill
]

\section{Introduction}

The increase in sensitivity of upcoming cosmic microwave background
(CMB) polarization experiments, both from ground and space, raises the possibility for detailed studies related to
 reionization. The main reason for this is the existence of a large angular scale polarization contribution
due to rescattering of the temperature quadrupole by free-electrons in the reionized epoch \cite{Zaldarriaga:1996ke}. 
This contribution peaks at angular scales corresponding to 
the horizon at the rescattering surface. While such a signal has now been detected with the temperature-polarization cross-correlation
power spectrum measured with first-year data of the Wilkinson Microwave Anisotropy Probe (WMAP) \cite{Bennett:2003bz}, information related to
reionization from it, unfortunately, is limited.  While the amplitude of the signal depends on the total optical
depth to electron scattering, with an estimated value of 0.17 $\pm$ 0.04 \cite{Kogetal03}, the reionization history 
related to this optical depth is still unknown. While  slight modifications to the large angular scale polarization
power spectra exist with complex reionization histories, such as a two stage reionization process advocated by Ref.~\cite{Cen:2002zc},
due to large cosmic variance associated with anisotropy measurements at few tens degree scales, one cannot fully reconstruct the
reionization history as a function of redshift \cite{Kaplinghat:2002vt}. 

Given the importance of understanding reionization for both cosmological and astrophysical reasons, and the
increase in sensitivity and angular resolution of upcoming CMB polarization data, it is useful to consider
alternative possibilities for additional studies beyond the large angular scale signal.
Under standard expectations for reionization, mainly due to UV light emitted by first luminous objects,
the process of reionization is expected to be patchy and inhomogeneous \cite{Barkana:2000fd}. This leads to fluctuations in the
electron scattering optical depth and a modulation to the polarization contribution such that new anisotropy fluctuations, at arcminute scales
corresponding to inhomogeneities in the visibility function, is generated \cite{Hu:1999vq,Santos:2003jb}. Even if the reionization were to 
be more uniform,
as expected from alternative models involving X-ray backgrounds \cite{Oh:2000zx} and reionization via particle decays \cite{Hansen:2003yj},
density fluctuations in the electron field can modulate the polarization contribution and generate secondary
anisotropies \cite{Baumann:2002es}. The new secondary polarization fluctuations,
whether due to patchiness or density inhomogeneities, however, is hardly detectable in the polarization power spectrum
given the dominant primordial polarization contribution, with a peak at arcminute scales, related to the velocity field at the
last scattering surface. Thus, information related to reionization is limited from measurements that involve simply the
two point correlation function of polarization or the cross-correlation between polarization and the temperature.
To extract additional information one must move to a higher order and consider measurements related to, for example,
the three-point correlation function or the bispectrum \cite{Cooray:1999kg}.

In this paper, we consider a potentially interesting study that has the capability to allow detailed measurements related to the
reionization history. The proposed measurement involves the cross-correlation between CMB polarization maps and 
images of the high redshift universe around the era of reionization or its end. The existence of such a correlation arises
from the fact that the polarization contribution related to reionization is generated at the same epoch as that related to the
tracer field selected from high redshifts. Since the reionization contribution to polarization is generated by the 
scattering of the temperature quadrupole, direct cross-correlation between polarization maps and a tracer field
is not useful. However, since the rescattering quadrupole correlates with CMB temperature maps,  
one must consider higher order statistics that also use information from CMB temperature data. We consider a
possibility in the form of a three point correlation function, or a bispectrum in Fourier space,
 involving CMB polarization maps at arcminute-scale resolution, CMB temperature maps with large angular scale
anisotropy information, and maps of the high redshift universe that trace fluctuations during the reionization and prior to reionization.
Examples of useful fields include maps of the infrared background anisotropies related to the redshifted UV emission 
by first proto galaxies \cite{Cooray:2003yf} and fluctuations in the 21 cm background due to the neutral Hydrogen content \cite{Tozzi:1999zh}. 
Additional possibilities also include maps of the $z \sim 3$ universe since one can then establish the low-redshift part of the 
electron scattering
visibility function. Cross-correlation studies with the 21 cm background may be the most interesting
possibility 
given that one is allowed to a priori select redshift ranges from which fluctuations arise to the redshifted 21 cm emission, 
based on frequency information of the observations.

The paper is organized as follows.  In \S~\ref{sec:deriv}, we derive the existence of a higher order correlation between
CMB temperature, polarization and a tracer field of the large scale structure that overlaps with fluctuations in the visibility function.
In \S~\ref{sec:results}, we discuss our results and suggest that there is adequate signal-to-noise to perform such a
measurements and consider several applications. We conclude with a summary in \S~\ref{sec:summary}.

\section{Calculation Method}
\label{sec:deriv}

The rescattering of CMB photons by free-electrons at redshifts significantly 
below decoupling, at $z \sim 1100$, leads to a polarized intensity, which in terms of the linear Stokes parameters, is
\begin{eqnarray}
_{\pm} X (\bn) &\equiv& \nonumber \\
 (q \pm iu)(\bn) &=& \frac{\sqrt{24 \pi}}{10} \int d\rad g(\rad) \sum_{m=-2}^{2} Q^{(m)}\;  \Ytwotwomn(\bn) \, ,
\label{eqn:X}
\end{eqnarray}
where the quadrupole anisotropy of the radiation field is
\begin{equation}
Q^{(m)}(\bx) = - \int d\Omega \frac{\Ytwomn}{\sqrt{4\pi}} \Theta(\bx,\bn) \, .
\end{equation}
In Eq.~\ref{eqn:X}, $r(z)$ is the conformal distance (or lookback time) from the observer at redshift $z=0$, given by
\begin{equation}
\rad(z) = \int_0^z {dz' \over H(z')} \,,
\end{equation}
where the expansion rate for adiabatic CDM cosmological models with a cosmological constant is
\begin{equation}
H^2 = H_0^2 \left[ \Omega_m(1+z)^3 + \Omega_K (1+z)^2
              +\Omega_\Lambda \right]\,,
\end{equation}
where $H_0$ can be written as the inverse
Hubble distance today $H_0^{-1} = 2997.9h^{-1} $Mpc.
We follow the conventions that
in units of the critical density $3H_0^2/8\pi G$,
the contribution of each component is denoted $\Omega_i$,
$i=c$ for the CDM, $b$ for the baryons, $\Lambda$ for the cosmological
constant. We also define the
auxiliary quantities $\Omega_m=\Omega_c+\Omega_b$ and
$\Omega_K=1-\sum_i \Omega_i$, which represent the matter density and
the contribution of spatial curvature to the expansion rate
respectively. Although we maintain generality in all derivations, we
illustrate our results with the currently favored $\Lambda$CDM
cosmological model. The parameters for this model
are $\Omega_c=0.30$, $\Omega_b=0.05$, $\Omega_\Lambda=0.65$ and $h=0.65$.

The visibility function, or the probability of scattering within $d\rad$ of $\rad$, is
\begin{equation}
g =  \dot \tau e^{-\tau} = X_e(z) H_0 \tau_H (1+z)^2 e^{-\tau}\,.
\end{equation}
Here
$\tau(r) = \int_0^{\rad} d\rad \dot\tau$ is the optical depth out to $r$,
$X_e(z)$ is the ionization fraction, as a function of redshift, and
\begin{equation}
       \tau_H = 0.0691 (1-Y_p)\Omega_b h\,,
\end{equation}
is the optical depth to Thomson
scattering to the Hubble distance today, assuming full
hydrogen ionization with primordial helium fraction of $Y_p (=0.24)$.
In typical calculations of CMB anisotropies, the general assumption is that the universe reionized smoothly
and promptly such that $X_e(z \lesssim z_{ri})=1$ and $X_e(z > z_{ri})=0$. Such an assumption, however, is
in conflict with observations, so far, involving the optical depth to electron scattering measured by WMAP and the
presence of a 1\% neutral fraction based on Lyman-$\alpha$ optical depths related to Gunn-Peterson troughs of the
$z \sim 6$ quasars in the Sloan Digital Sky Survey \cite{Fan:2001vx}. To explain the WMAP optical depth and
low redshift data simultaneously require complex reionization histories where $X_e(z)$ is not abrupt, but where
reionization takes a long time \cite{Cen:2003ey}. Since first luminous objects are usually considered as a source of
reionization, the process is inhomogeneous and patchy. 

\begin{figure*}[!th]
\centerline{\psfig{file=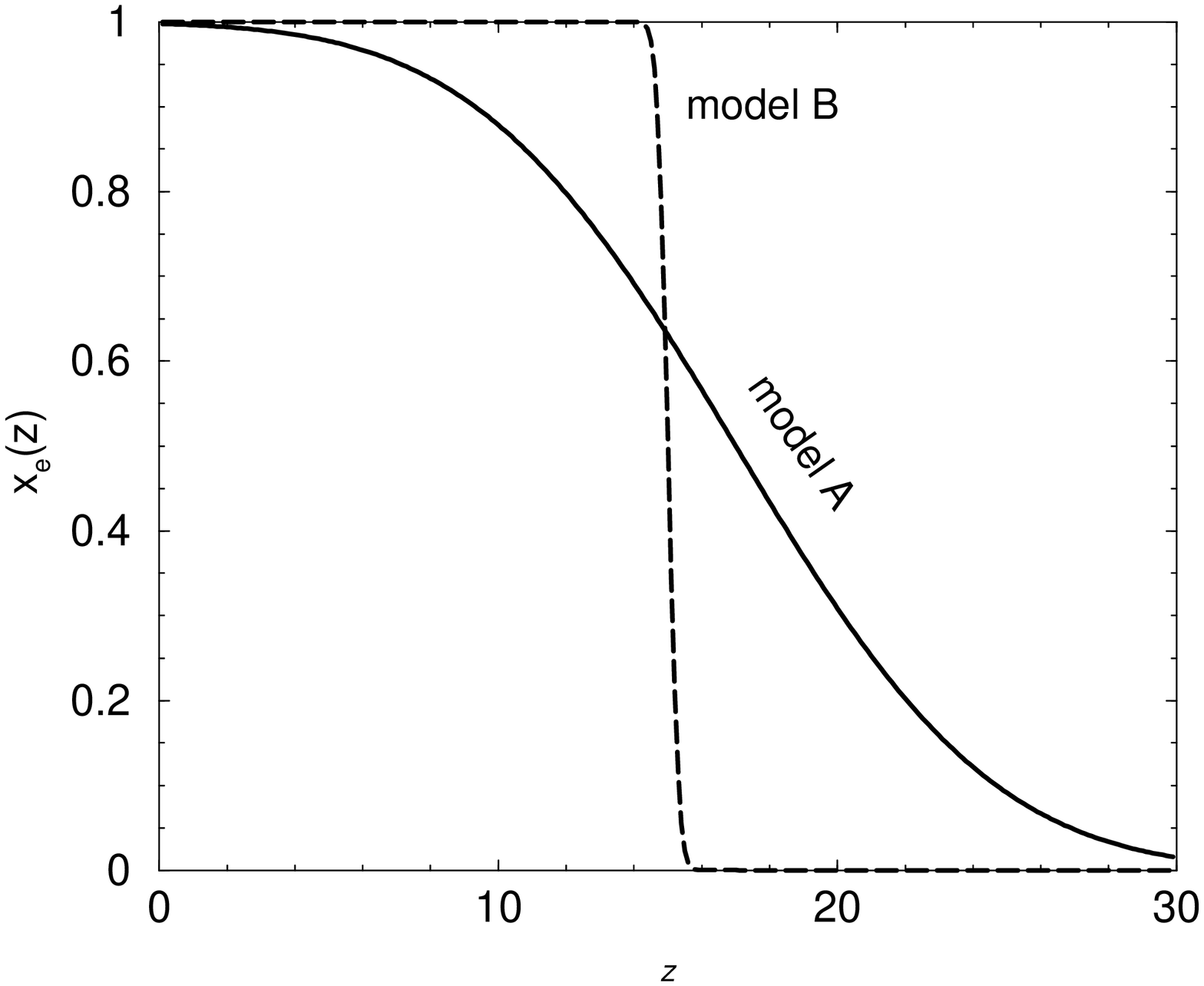,width=3.4in,angle=-90}
\psfig{file=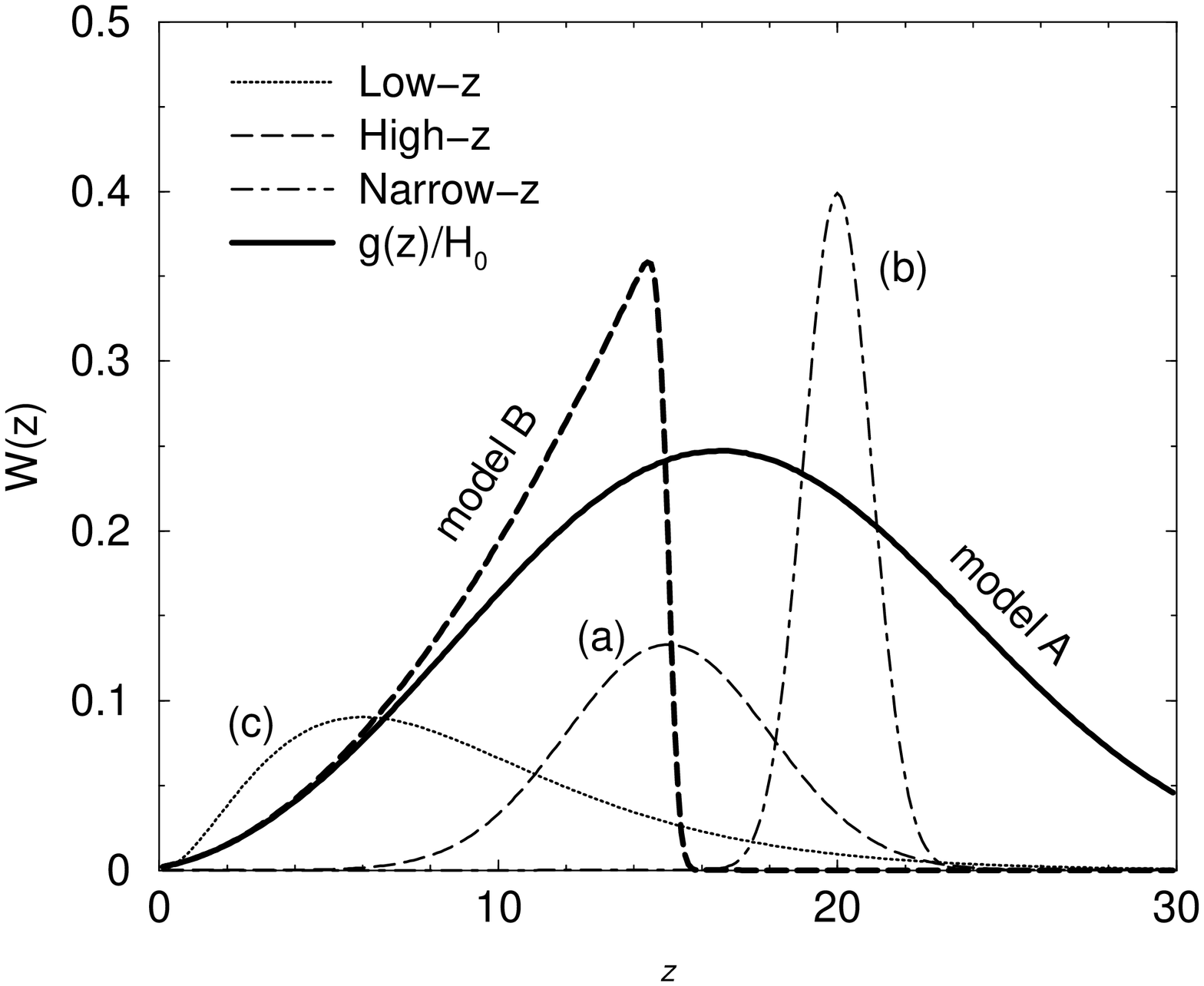,width=3.4in,angle=-90}}
\caption{{\it Left panel:} The reionization history of the universe. Here, we consider two possibilities for the reionization
process involving a smooth transition over a wide range of redshift (model A) and a model which involve a sudden transition to a
fully reionized model at a reionization redshift $\sim$ 15. 
{\it Right panel:} The visibility function related to electron scattering in the reionized epoch, $g(z)$ (see, equation~5) for both models A and B
shown above.  For comparison, we also show several normalized redshift distributions of tracer fields
that we consider here. These are, (a) a high redshift broad distribution, 
such as the one related to first proto-galaxy contributions to the IR background, 
(b) a high redshift, but narrow, distribution expected from the 21cm background fluctuation studies in the future, and
(c) a low redshift distribution, such as the distribution of $z \sim$ 3 galaxy population.
Note that 21 cm fluctuation studies allow one to choose narrow bands, such as (b), 
in redshift over a wide range both prior to and during the reionization  process
and may provide the optimal tracer field for the proposed study here.}
\label{fig:weight}
\end{figure*}

Here, we make use of two descriptions of $X_e(z)$ (Fig.~1; left panel). In the first case (model A), 
we use a calculation based on  the Press-Schechter \cite{Press:1973iz} mass function and related to the reionization by UV light from
the first-star formation following Ref.~\cite{Chen:2003sw}. Here, $X_e(z)$ varies from a value less than 10$^{-1}$ at a redshift of 30
to a value of unity when the universe is fully reionized, at a redshift of $\sim$ 5. The reionization history is rather broad
and the universe does not become fully reionized till late times, though the reionization process began at a much higher epoch.
The total optical depth related to this ionization history is $\sim$ 0.17, consistent with WMAP measurements \cite{Kogetal03}. 
As an alternative to such a history, and to consider the possibility how different reionization models can be studied with the
proposed cross-correlation analysis, we also consider a reionization history that is instantaneous at the redshift of reionization
of $\sim$ 15 (model B). This alternative model could arise from reionization descriptions that involve the presence of a X-ray background or
where the reionization process is associated with decaying particles, among others. While the transition to a fully reionized
universe, from a neutral one, is sudden, a history such as the one shown produces a similar optical depth to electron scattering as in
model A. 

In  Fig.~1 right panel, we show the visibility function $g(z)$ related to these two reionization histories. 
in model A, scattering is distributed widely given the long reionization process, while in the case of model B,  scattering is
mostly concentrated during the sharp transition to a reionized universe. As shown in Fig.~1, one does not expect
a probe of the high-z universe, say at a redshift $\sim$ 20, to correlate strongly with polarization if
the reionization process is more like model B, while a strong correlation is expected if the hypothesis related to a lengthy reionization
process does in fact happen.

Now we will discuss how these reionization histories determine polarization signals in CMB both at large and small angular scales.
First, we note that the polarization is a spin-2 field and can be decomposed using the spin-spherical harmonics \cite{Gol67}
such that
\begin{equation}
_{\pm} X (\bn) = \sum_{lm} \; _{\pm} X_{lm} \; \Ytwolmn(\bn) \, .
\label{eqn:multi}
\end{equation}
The multipole moments of polarization field are generally separated to ones with gradient ($E$) and curl ($B$) parity
such that \cite{Zaldarriaga:1996xe}
\begin{equation}
_{\pm} X_{lm} = E_{lm} \pm i B_{lm} \, .
\end{equation} 
Instead of Stokes-Q and -U, one can redefine two polarization related fields $E(\bn)$, a scalar, and $B(\bn)$, a pseudo-scalar, such that
\begin{eqnarray}
E(\bn)&=& \sum_{lm} E_{lm} \Ylmn(\bn) \, \nonumber \\
B(\bn)&=& \sum_{lm} B_{lm} \Ylmn(\bn) \, .
\end{eqnarray}

The cross correlation between reionization generated polarization and a tracer field of the large scale structure is zero.
This can be understood based on the fact that 
\begin{equation}
\langle _{\pm} X (\bn) S(\bn') \rangle \propto \sum_m \langle Q^{(m)} S(\bn') \rangle = 0 \, ,
\end{equation}
since the temperature quadrupole that rescatters at low redshifts is not correlated with
the local density field. There is, however, a non-zero cross-correlation between polarization and the CMB temperature
anisotropy due to the fact that the
temperature quadrupole related to scattering is present in the CMB temperature anisotropy map:
\begin{equation}
\langle _{\pm} X (\bn) \Theta(\bn') \rangle \propto \sum_m \langle Q^{(m)} \Theta(\bn') \rangle \neq 0 \, .
\end{equation}
Since this is part of the standard calculation
related to the $C_l^{TE}$ angular
power spectrum, we do not further consider it.

To the next highest order, contributions to cross-correlation arise from the fact that fluctuations in the
electron field is correlated with the large scale structure. This second-order term arises from the fact that
\begin{equation}
g(\bn,\rad) = \bar{g}(\bn,\rad) \left[ 1 + \frac{\delta g(\bn,\rad)}{\bar{g}(\bn,\rad)}\right] \, .
\label{eqn:g}
\end{equation}
These fluctuations lead to a second order  polarization signal that peaks at arcminute angular scales
and are discussed in Ref.~\cite{Hu:1999vq,Baumann:2002es} 

Substituting Eq.~\ref{eqn:g} in Eq.~\ref{eqn:X} and using Eq.~\ref{eqn:multi}, we can write the multipole moments the polarization field,
in the presence of fluctuations in the visibility function, as
\begin{eqnarray}
&& _{\pm} X_{lm}  = -\frac{\sqrt{24} \pi}{5\sqrt{5}} (4\pi)^2 \sum_{l_1 m_1}\sum_{l_2 m_2} \sum_{m_3} i^{l_1+l_2} \nonumber \\
&\times&\int d\rad \bar{g} \int \frac{d^3{\bf k_1}}{(2\pi)^3} \int \frac{d^3{\bf k_2}}{(2\pi)^3} Q^{(0)}(k_1) 
\delta_g(k_2) j_{l_2}(k_2\rad) j_{l_1}(k_1 \rad)\nonumber \\
&\times& \Ylm1(\veck_1) \Ylm2(\veck_2) \Ytwotwom3(\veck_1) \nonumber \\
&\times& \int d\Omega\; \Ytwolmn(\bn) \Ylm1(\bn) \Ylm2(\bn) \Ytwotwom3\,^*(\bn) \, ,
\end{eqnarray}
where, following Ref.~\cite{Hu:1999vq}, we have projected the fluctuations-modulated temperature quadrupole
\begin{equation}
Q_g^{(m)}(\bk) = \int \frac{d^3{\bf k_1}}{(2\pi)^3} Q^{(m)}(\bk_1) \delta_g (|\bk-\bk_1|) \, ,
\end{equation}
to a basis where the $z$-axis is parallel to the $\bk$ vector such that
\begin{equation}
Q_g^{(m)}(\bk)= \sqrt{\frac{4\pi}{5}} \int \frac{d^3{\bf k_1}}{(2\pi)^3} Q^{(0)}(\bk_1) \delta_g (\bk_2)  \Ytwomn(\veck_1) \, ,
\end{equation}
and have simplified using the Rayleigh expansion of a plane wave
\begin{equation}
e^{i{\bf k}\cdot \hat{\bf n}\rad}=
4\pi\sum_{lm}i^lj_l(k\rad)Y_l^{m \ast}(\bk)
\Ylmn(\bn)\, .
\label{eqn:Rayleigh}
\end{equation}

Since what correlates with the polarization field is the quadrupole generated by
large angular scale temperature fluctuations, to simplify, we make the assumption that the relevant anisotropies 
are due to the Sachs-Wolfe (SW; \cite{Sachs:er}) effect
\begin{equation}
\Theta(\bx,\bn) = - \frac{1}{3} \Phi(\bx,\bn)\Big|_{\rad=\rad_0}\, ,
\label{eq:sw}
\end{equation}
where $\rad_0 \equiv \rad(z=1100)$.
Note that there is a correction here related to the integrated Sachs-Wolfe (ISW) effect, 
but since it is present only at redshifts less than 1, when the dark energy component begins
to dominate the energy density of the universe, we ignore its contribution to the quadrupole. 
This is a valid assumption for models of reionization with an optical depth at
the level of 0.1 or more since most of the rescattering is then restricted to redshifts greater than 10 or more\footnote{When 
calculating the signal-to-noise, however,
we include all contributions to anisotropies, such that the temperature anisotropy angular power spectrum, $C_l^{\rm CMB}$, is not just due to
the SW effect.}.

Our assumption related to large angular scale fluctuations allows us to write the quadrupole at late times, when
projected to an observer at a distance $\rad$ through free-streaming, as
\begin{equation}
Q^{(0)}(\bk,\rad) = -\sqrt{5} \frac{1}{3} \Phi(\bk,\rad_0) j_2(\bk \rad_s) \, ,
\end{equation}
where $\rad_s = \rad_0-\rad$.

Using Eq.~\ref{eq:sw}, multipole moments of the temperature map at large angular scales are
\begin{equation}
\Theta_{lm} = - \frac{4\pi}{3} i^{l} \int \Dk \Phi(\bk,\rad_0) j_{l}(k\rad_0) \Ylmn(\veck) \, .
\end{equation}
Similarly, multipole moments of the tracer field are
\begin{equation}
S_{lm} = \int d \Omega \, \Ylmn\,^*(\bn)\, S(\bn) \, ,
\end{equation}
and assuming that the tracer field can be described with a source radial distribution of $W^S(\rad)$ with fluctuations denoted 
by $\delta_S$, we simplify to write
\begin{equation}
S_{lm} = 4\pi i^{l} \int \Dk \int d\rad W^S(\rad) \delta_S(\bk,\rad) j_{l}(k\rad) \Ylmn(\veck) \, .
\end{equation}

The cross-correlation with the large scale structure exists in the form of a bispectrum between
polarization-temperature-tracer fields. For this, we construct, for example,
$\langle E_{l_1 m_1} \Theta_{l_2 m_2} S_{l_3 m_3} \rangle$ and after some straightforward but tedious algebra,
we write
\begin{eqnarray}
&&\langle E_{l_1 m_1} \Theta_{l_2 m_2} S_{l_3 m_3} \rangle = \nonumber \\
&&4\pi\int d\rad_1 \int d\rad_2  I_{l_3}^{gS}(\rad_1,\rad_2) J_{l_2}^{\Phi}(\rad_1) \nonumber \\
&\times& \int d\Omega \Ylm1(\bn) \Ylm2(\bn) \Ylm3(\bn) \, ,
\end{eqnarray}
where
\begin{eqnarray}
I_{l}^{gS}(\rad_1,\rad_2) &=& \frac{2}{\pi}\int k^2 dk  g(\rad_1) W^S(\rad_2) j_{l_3}(k \rad_2) j_{l_3}(k\rad_1) P_{gS}(k) \, , \nonumber \\
J_{l}^{\Phi}(\rad) &=& \int \frac{k^2 dk}{18\pi^2}  P_{\Phi\Phi}(k_1,\rad_0) j_{l}(k \rad_0) j_2(k \rad_s) \epsilon_l(k\rad)\, .
\label{eqn:Ils}
\end{eqnarray}
Here, $\epsilon_l(x) = -j_l(x) + j''_k(x) + 2 j_l(x)/x^2 + 4 j_l'(x)/x$ and
we have defined three-dimensional power spectra related to the potential-field at the last scattering surface and  the cross-power
spectrum between fluctuations in the scattering visibility function and the tracer field:
\begin{eqnarray}
\left< \Phi({\bf k})\Phi({\bf k')} \right> &=& (2\pi)^3
        \deld({\bf k}+{\bf k'}) P_{\Phi\Phi}(k) \nonumber \\
\left< \delta_g({\bf k})\delta_S({\bf k')} \right> &=& (2\pi)^3
        \deld({\bf k}+{\bf k'}) P_{gS}(k) \, ,\nonumber \\
\end{eqnarray}
respectively, where $\deld$ is the Dirac delta function. 

In these calculations, we will described
$P_{gS}(k,z)$ following the halo model \cite{Cooray:2002di}. At large angular scales, relevant to most of the estimates
in here, $P_{gS}(k,z) = b_g(z) b_S(z) G^2(z) P^\lin(k)$, where $P^\lin(k)$ is the power spectrum of density fluctuations 
in linear perturbation theory, and $b_g(z)$ and $b_S(z)$ are large physical scale biases of these 
two fields taken to be independent of the scale 
(see, below).   We use the fitting formulae of Ref.~\cite{Eisenstein:1997jh} in evaluating the
transfer function for CDM models, while we adopt the COBE normalization to normalize the
linear power spectrum \cite{Bunn:1996da}. This model has mass fluctuations on the $8 h$ Mpc$^{-1}$
scale in accord with the abundance of galaxy clusters such that $\sigma_8=0.86$.  

The bispectrum for the case with E-mode map is given by
\begin{eqnarray}
&&B^{E\Theta S}_{l_1 l_2 l_3} = \sum_{m_1 m_2 m_3} \wjm \left< E_{l_1 m_1} \Theta_{l_2 m_2} S_{l_3 m_3}  \right> \nonumber \\
&=&\sqrt{\frac{(2l_1 +1)(2 l_2+1)(2l_3+1)}{4 \pi}} \left(
\begin{array}{ccc}
l_1 & l_2 & l_3 \\
0 & 0  &  0
\end{array}
\right) b^E_{l_2,l_3} \, , \nonumber \\
\label{eqn:ovbidefn}
\end{eqnarray}
where, 
\begin{eqnarray}
b^E_{l_2,l_3} &=&  4\pi
\int d\rad_1 \int d\rad_2  J_{l_2}^{\Phi}(\rad_1) I_{l_3}^{gS}(\rad_1,\rad_2) \, . \nonumber \\
\label{eqn:b}
\end{eqnarray}

One can simplify Eq.~\ref{eqn:b} with the use of the Limber approximation. Here, we use a version based on
the completeness relation of spherical Bessel functions \cite{Cooray:1999kg}
\begin{equation}
\int dk k^2 F(k) j_l(kr) j_l(kr')  \approx {\pi \over 2} \da^{-2} \deld(r-r')
                                                F(k)\big|_{k={l\over d_A}}\,,
\end{equation}
where the assumption is that $F(k)$ is a slowly-varying function
and the angular diameter distance in above is
\begin{equation}
\da = H_0^{-1} \Omega_K^{-1/2} \sinh (H_0 \Omega_K^{1/2} \rad)\,.
\end{equation}
Note that as $\Omega_K \rightarrow 0$, $\da \rightarrow \rad$.

\begin{figure}[!h]
\centerline{\psfig{file=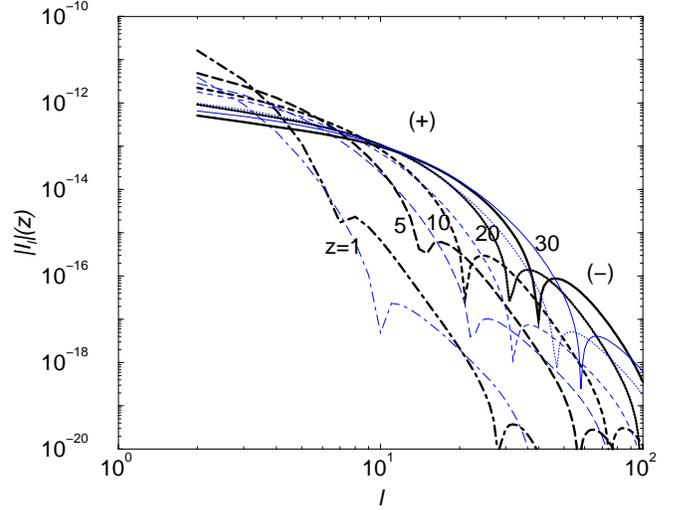,width=3.4in,angle=-90}}
\caption{The integrals $I_{l}^{E}(\rad)$, with thick lines (Eq.~\ref{eqn:Il}), and $I_{l}^{B}(\rad)$, with thin lines (Eq.~\ref{eqn:IlB}), 
as a function of the multipole for several different redshift values.  These mode coupling integrals 
oscillate between positive and negative values and we label
 positive/negative parts with (+)/(-). These functions captures the cross-correlation between
density modulated temperature quadrupole that rescatter by an electron, at a radial distance of $\rad$,  
to form E- and B-modes of polarization and the temperature anisotropies seen by an
observer at a redshift of zero or today. The multipole $l$ here is that of  temperature anisotropies as seen today.} 
\label{fig:integral}
\end{figure}

Applying this approximation to the integral in $I_{l_3}^{gS}$ (Eq.~\ref{eqn:Ils}) allows us to further simplify and write
\begin{eqnarray}
b^E_{l_2,l_3} &=&   \frac{2}{9\pi}
\int d\rad g \frac{G^2(\rad)}{\da^2} W^S(\rad) P_{gS}\left(k=\frac{l_3}{\da}\right)I_{l_2}^{E}(\rad) \, ,
\label{eqn:bl}
\end{eqnarray}
where 
\begin{eqnarray}
I_{l}^{E}(\rad) &=& \int k^2 dk  P_{\Phi\Phi}(k_1,\rad_0) j_{l}(k \rad_0) j_2(k \rad_s)   \epsilon_l(k \rad) \, . \nonumber \\
\label{eqn:Il}
\end{eqnarray}

Similar to the bispectrum involving the E-mode map, combined with temperature and a tracer field of the high-redshift universe,
one can also construct a bispectrum involving the B-mode map  instead of the E-mode map. As discussed in Ref.~\cite{Hu:1999vq},
rescattering under density fluctuations to the electron distribution also lead to a contribution to the B-mode map
and can be understood based on the fact that the density-modulated quadrupole generates $m=\pm 1$ components of the quadrupole
which when scattered lead to B-modes\footnote{Without density fluctuations, the temperature quadrupole related to scalar fluctuations
induces only the $m=0$ component and only E-modes of the polarization are generated at large angular scales.}.
Following our earlier discussion, this bispectrum is written as
\begin{eqnarray}
&&B^{B\Theta S}_{l_1 l_2 l_3} = \sum_{m_1 m_2 m_3} \wjm \left< B_{l_1 m_1} \Theta_{l_2 m_2} S_{l_3 m_3}  \right> \nonumber \\
&=&\sqrt{\frac{(2l_1 +1)(2 l_2+1)(2l_3+1)}{4 \pi}} \left(
\begin{array}{ccc}
l_1 & l_2 & l_3 \\
0 & 0  &  0
\end{array}
\right) b^B_{l_2,l_3} \, , \nonumber \\
\label{eqn:ovbiBdefn}
\end{eqnarray}
where, 
\begin{eqnarray}
b^B_{l_2,l_3} &=&   \frac{2}{9\pi}
\int d\rad g \frac{G^2(\rad)}{\da^2} W^S(\rad) P_{gS}\left(k=\frac{l_3}{\da}\right)I_{l_2}^{B}(\rad) \, ,
\label{eqn:bl2}
\end{eqnarray}
where 
\begin{eqnarray}
I_{l}^{B}(\rad) &=& \int k^2 dk  P_{\Phi\Phi}(k_1,\rad_0) j_{l}(k \rad_0) j_2(k \rad_s)   \beta_l(k\rad) \, ,
\label{eqn:IlB}
\end{eqnarray}
where, the function that replaces $\epsilon_l$ in $I_{l}^{E}$ to obtain $I_{l}^{B}$ is $\beta_l(x) = 
2j_l'(x) + 4 j_l(x)/x$.

In Fig.~2, we  show the two quantities $I_{l}^{E,B}$ at several different values of $z$, as a function of $l$. While the
function peaks at multipoles less than 4, at $z=1$, the function broadens 
as one moves to redshifts greater than 10 such that it is basically a constant for $l$ values up to $\sim$ 20 or more.
The quantities $I_{l}^{E,B}(\rad)$ only depends on the background 
cosmology and the primordial power spectrum;
with temperature anisotropy measurements and other cosmological studies, these parameters can assumed to be known.
Thus, the only unknown quantity will be related to reionization through the visibility function, $g(\rad)$.  
Thus, the cross-correlation between
tracer fields allow some information related to $g(\rad)$ be established. This includes not only whether $g(\rad)$ is
inhomogeneous (through the detection of fluctuations), but also 
information on the form of $g(\rad)$, as a function of $\rad$ or binned $\rad$. 

\begin{figure}[!h]
\centerline{\psfig{file=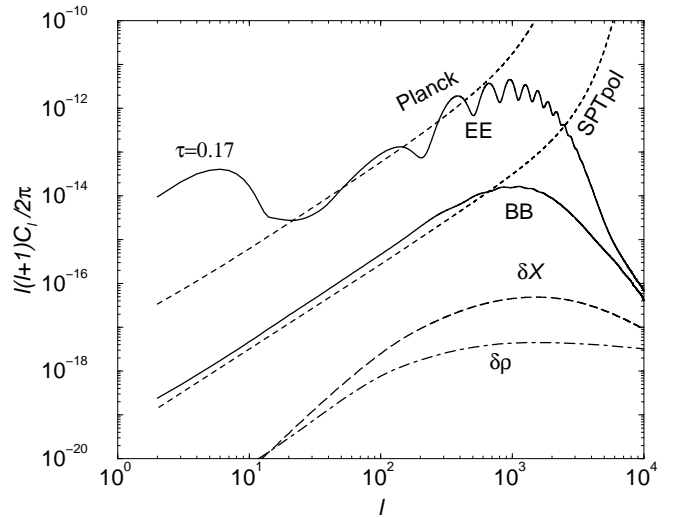,width=3.4in,angle=-90}}
\caption{The angular power spectrum of CMB polarization in both the dominant gradient- or the E-mode and the B-mode. 
In the case of E-modes, we show the primordial contribution
with the correction at large angular scales related to reionization scattering, and assuming a total optical depth to scattering of
0.17. In the case of B-modes, we neglect the primordial contribution related to gravity waves --- both due to the unknown
amplitude and the fact that this contribution peaks at large angular scales while the proposed correlation is related to arcminute scales ---
and show the important arcminute-scale signal related to lensing conversion of E- to B-modes.
The long-dashed and dot-dashed curves are the secondary polarization contributions, with power in E-modes equal to B-modes, 
related to inhomogeneities in the reionization
involving ionized fraction  and the electron density field, respectively. For reionization histories related to the formation of
first luminous objects, the reionization is significantly inhomogeneous and leads to a higher secondary contribution related to
patchiness of the reionization when compared to fluctuations in the electron density. The two contributions also are generated at
different epochs, with the patchy contribution resulting from redshifts where $0 <X_e(z) < 1$, while the density fluctuations generate
contributions even when $X_e(z)=1$.}
\label{fig:cl}
\end{figure}

\begin{figure}[!th]
\centerline{\psfig{file=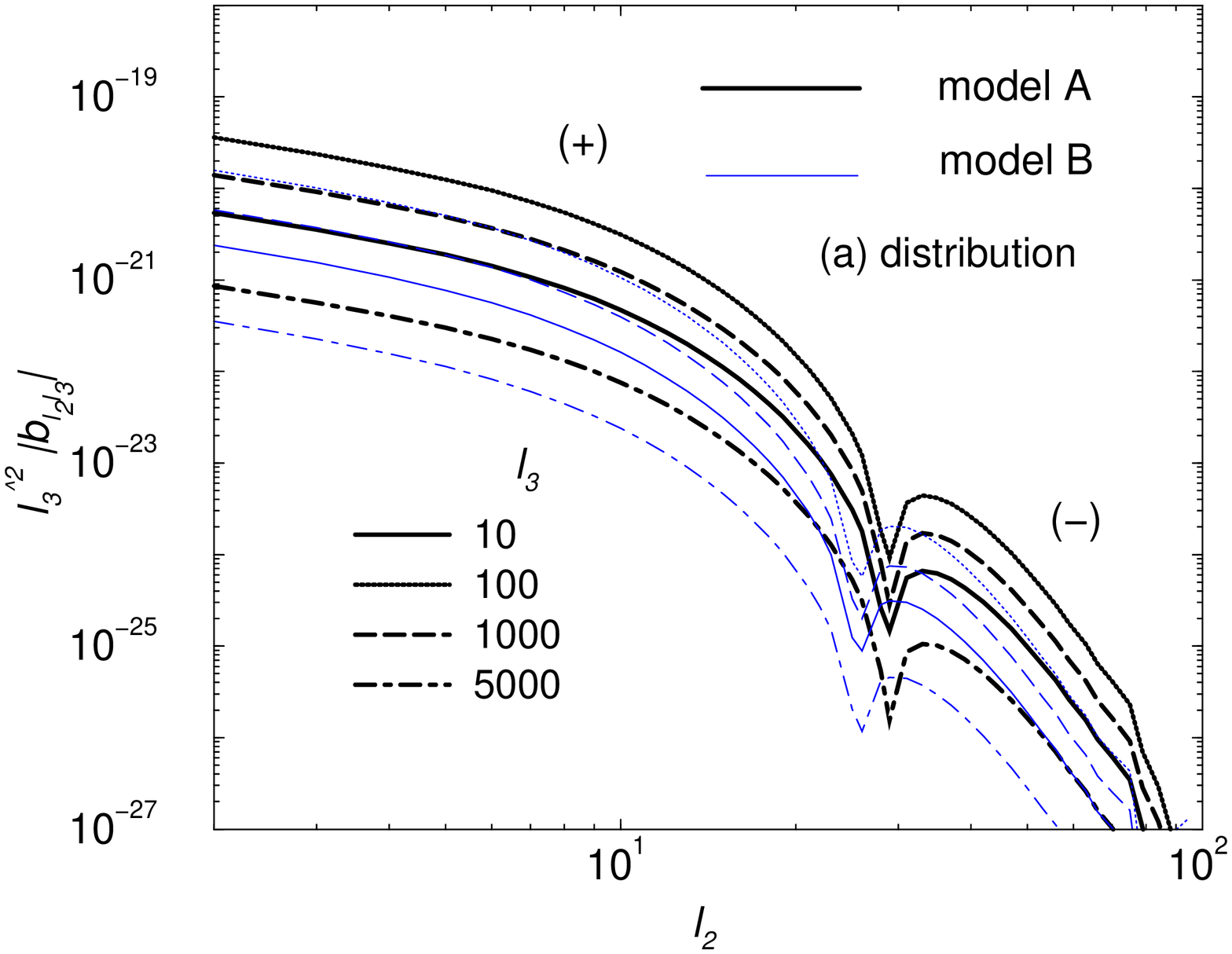,width=3.4in,angle=-90}}
\centerline{\psfig{file=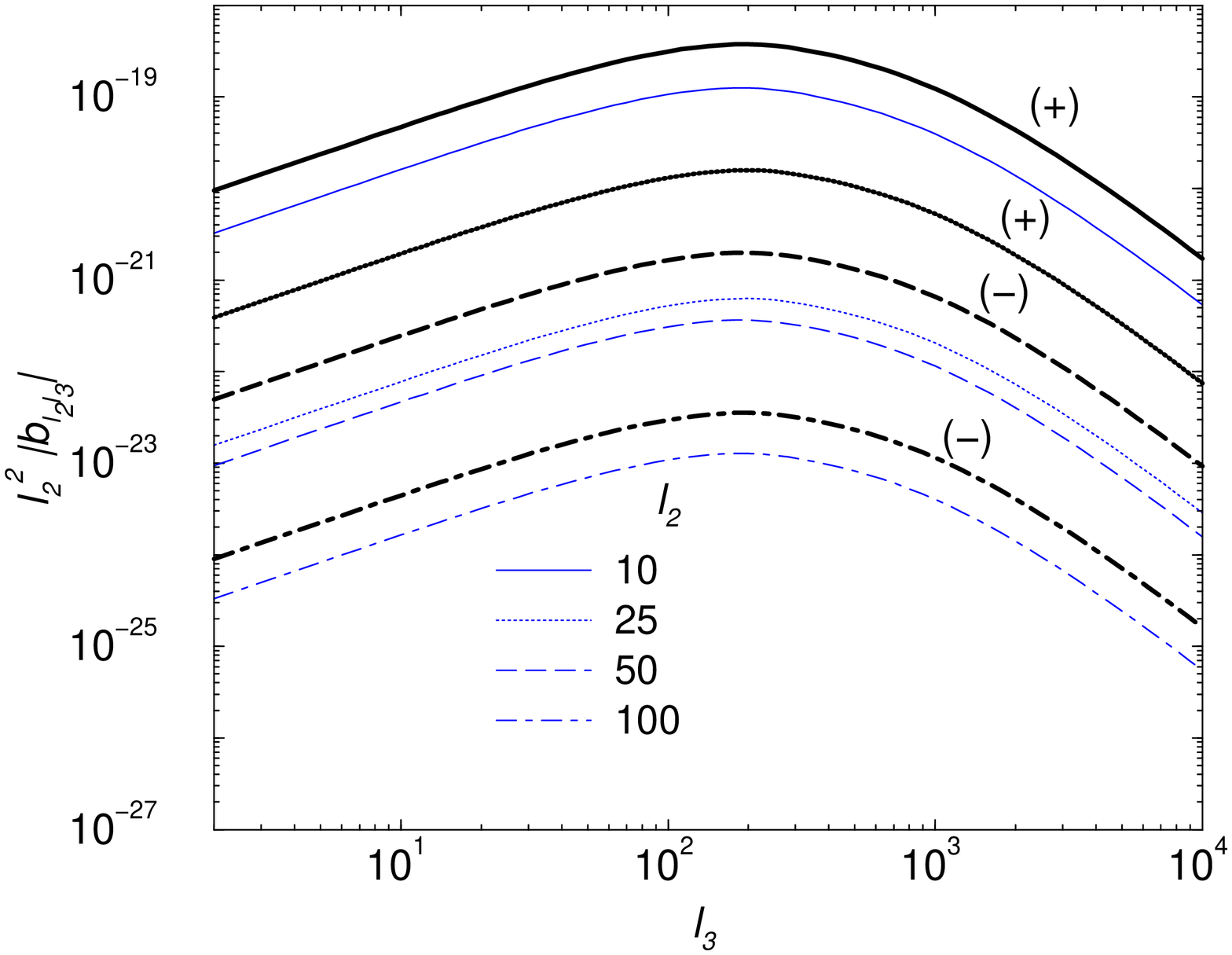,width=3.4in,angle=-90}}
\caption{$b_{l_2,l_3}$ as a function of the mutlipole for the two reionization models separated to thick (model A) and thin (model B) lines.
In the top panel, we show $b_{l_2,l_3}$ as a function of $l_2$ for several different values of $l_3$ while in the bottom panel we show
$b_{l_2,l_3}$ as a function of $l_3$ for several values of $l_2$. The curves labeled (+)/(-) are positive/negative values of this quantity.
Since $b_{l_2,l_3} \propto \int dr f(l_3) I_{l_2}$ the top panel is
simply reflects $I_{l_2}$, as shown in Fig.~2. Here, we assume a tracer field with a distribution given by the broad high-z distribution 
labeled (a) in Fig.~1. As shown in the bottom panel, $b_{l_2,l_3}$ values, as a function of $l_3$, for the two different reionization models,
cannot be simply scaled by an overall normalization. Since $I_{l_2}$ is summed over the reionization visibility function, this is merely due to
the difference in the reionization duration. The same can be inferred through Fig.~2 where, as the redshift is increased, $I_{l}$ curves
shift to higher multipoles.}
\label{fig:bl}
\end{figure}

\section{Results}
\label{sec:results}

\subsection{Cross-correlation}

It is now well known that the large scale temperature fluctuations correlate with fluctuations in polarization
generated by rescattering due to free-electrons in the reionized epoch.
Though there is no direct cross-correlation between polarization data and maps of the large scale structure,
we suggest that the correlation between fluctuations in the large scale structure density fields and  that of the
electron scattering visibility function can be extracted by constructing a three-point correlation function, or a bispectrum
in Fourier space, involving the large angular scale CMB temperature anisotropies, arcminute scale polarization fluctuations,
and maps of the high redshift universe. The requirements for the existence of such a correlation are (1) the presence of
fluctuations in the visibility function, due to density inhomogeneities or patchiness of the reionization process, and (2)
overlap between the redshift distribution of scattering electrons, or the mean visibility function, and that of the tracer field. 

In Fig.~1, we illustrate the basis for the proposed study. Here, we show two visibility functions related
electron scattering based on two different descriptions of reionization history consistent with current estimates of the optical depth;
In addition to these two, note that one can conceive a large number of reionization history models that given the same optical depth \cite{Cen:2003ey}.
In the first scenario involving reionization by the UV light from first stars (model A),
 the visibility function is rather broad, but generally peaks at redshifts where $X_e(z) \sim 0.5$, while in the second scenario
with a  sudden transition to a reionized universe, scattering happens mostly during this transition.

To describe the three-dimensional cross power spectrum between the tracer field and that of the visibility function, we make use of
the halo model, but concentrate only on large angular scale clustering captured by the 2-halo term. In this limit, the relevant bias
factors can be described as
\begin{equation}
b_i(z) =  \frac{\int_{M_-}^{M_+} dM\, M\, \frac{dN}{dM} b(M,z)}{\int_{M_-}^{M_+} dM\, M\, \frac{dN}{dM}} \, ,
\label{eqn:bias}
\end{equation}
where $M_-$ and $M_+$ are the lower and upper limits of masses and $b(M,z)$ is the halo bias. We consider two
possibilities for fluctuations in $g(\rad)$: (1) density inhomogeneities and (2) patchiness. The bias factor
related to the patchy model follows from Ref.~\cite{Santos:2003jb}, while in the case of density inhomogeneities, we consider
a constant bias factor of unity. This allows us to write $b_g(z)$ as
\begin{equation}
b_g(z) \approx 1.0 + b_{X_e}(z) \, ,
\end{equation}
where $b_{X_e}(z) = 0$ when $X_e(z) = 1$, such that we no longer consider patchiness of the reionization when the universe is fully
reionized. The bias factor related to $X_e$ fluctuations are described in Ref.~\cite{Santos:2003jb} to which we refer the reader for further details;
an interesting point related to this bias factor is that it is dominated by halos with temperature at the level of $10^4$ K  and above,
where atomic cooling is expected and first objects form and subsequently reionize the universe. Since what enters in the cross-correlation is
the bias factor times the growth of density perturbations, in this case, the product $b_{X_e}(z)G(z)$ 
is in fact a constant \cite{Pee80,Oh:2003sa}, as a function of redshift out to z of 20 or more
given the rareness of halos at high redshifts.

The source bias $b_S(z)$, for the three tracer field distributions in Fig.~1
again involves a similar description. In the case of our low redshift distribution --- curve labeled (c) in Fig.~1--- we take $b_S(z)=1$, to
be consistent with typical bias involved with galaxy fields. For the two high redshift ones, we calculate the
bias following Eq.~\ref{eqn:bias}. For the broad high-z distribution, we take a bias factor consistent with the formation of
first objects and set $M_-$ to be the value corresponding to virial temperatures of 10$^4$K and $M_+ \rightarrow \infty$.
For the high-z narrow distribution, we assume that neutral gas is only present in halos other than those that form 
first objects (and, thus, assumed to be reionized). Thus we take $M_- \rightarrow 0$ and $M_+$ the value corresponding to virial
temperatures of 10$^4$K; note that we have taken a simple description of source bias here for both halos containing first
luminous objects and halos containing neutral Hydrogen. The situation is likely to be more complicated, but our main objective
is to show that there is adequate signal-to-noise for a detection of the proposed polarization-temperature-high redshift cross-correlation.

For models of extended reionization, the patchiness is more important. We illustrate this in Fig.~3 in terms of polarization power
spectra.  For reference, we show the primordial polarization anisotropy spectra generated at the last scattering, including the homogeneous rescattering
contribution at large angular scales. The secondary scattering contributions, due to inhomogeneities, can be written as \cite{Hu:1999vq,Baumann:2002es} 
\begin{equation}
C_l = \frac{3}{100} \int d\rad \frac{g^2(\rad)}{\da^2} Q_{\rm rms}^2(\rad) P_{gg}\left(k=\frac{l}{\da},\rad\right)\, ,
\end{equation}
where $Q_{\rm rms}^2(\rad)= \int k^2 dk/2\pi^2 Q^{(0)}(k,\rad)$. Note that the patchy polarization is generally higher \cite{Santos:2003jb} and
may be the most important secondary polarization contribution related to scattering at arcminute scales. 

Fig.~3 also illustrates why the secondary polarization contributions may be undetectable from the 
power spectrum alone; since the peak of the secondary contribution
is at the same angular scales as the primordial power spectrum peak, with an amplitude four-orders of magnitude smaller, 
the cosmic variance may not allow one to extract the secondary polarization contribution easily. For comparison, in Fig.~3,
we also show the expected noise power spectra related to two upcoming measurements of polarization involving
Planck from space and a ground-based experiment, such as using a large-format polarization-sensitive bolometer array, 
with 1000 pixels, 
at the South Pole Telescope. We take one arcminute resolution, but limited to a sky-coverage of 4000 deg.$^2$.  Other possibilities 
include experiments such as QUAD \cite{Bowden:2003ub}, but with slightly lower resolution (at the level of 5 arcminutes).

Given that arcminute scale polarization information is correlated with the tracer field, one does not require
all-sky maps of polarization or that of the tracer field. 
The required temperature information, however, comes from large angular scales, as we have shown in Fig.~2 with the integral of
$I_l^\Phi$. The $l$ value here corresponds to the temperature anisotropy multipole and what correlates with polarization
fluctuations is the large angular scale temperature fluctuations. On the other hand, at the same time,
arcminute-scale density fluctuations correlate with fluctuations in the visibility function and those responsible for secondary
polarization signal. The large angular scale temperature information, fortunately,
is already available from the WMAP data to the limit allowed by the cosmic variance. 
While CMB polarization information, at arcminute scales, will soon be avilable both from ground and space, 
the extent to which the proposed study can be carried out will be limited from
maps of the high redshift universe. 

\begin{figure*}[!th]
\centerline{\psfig{file=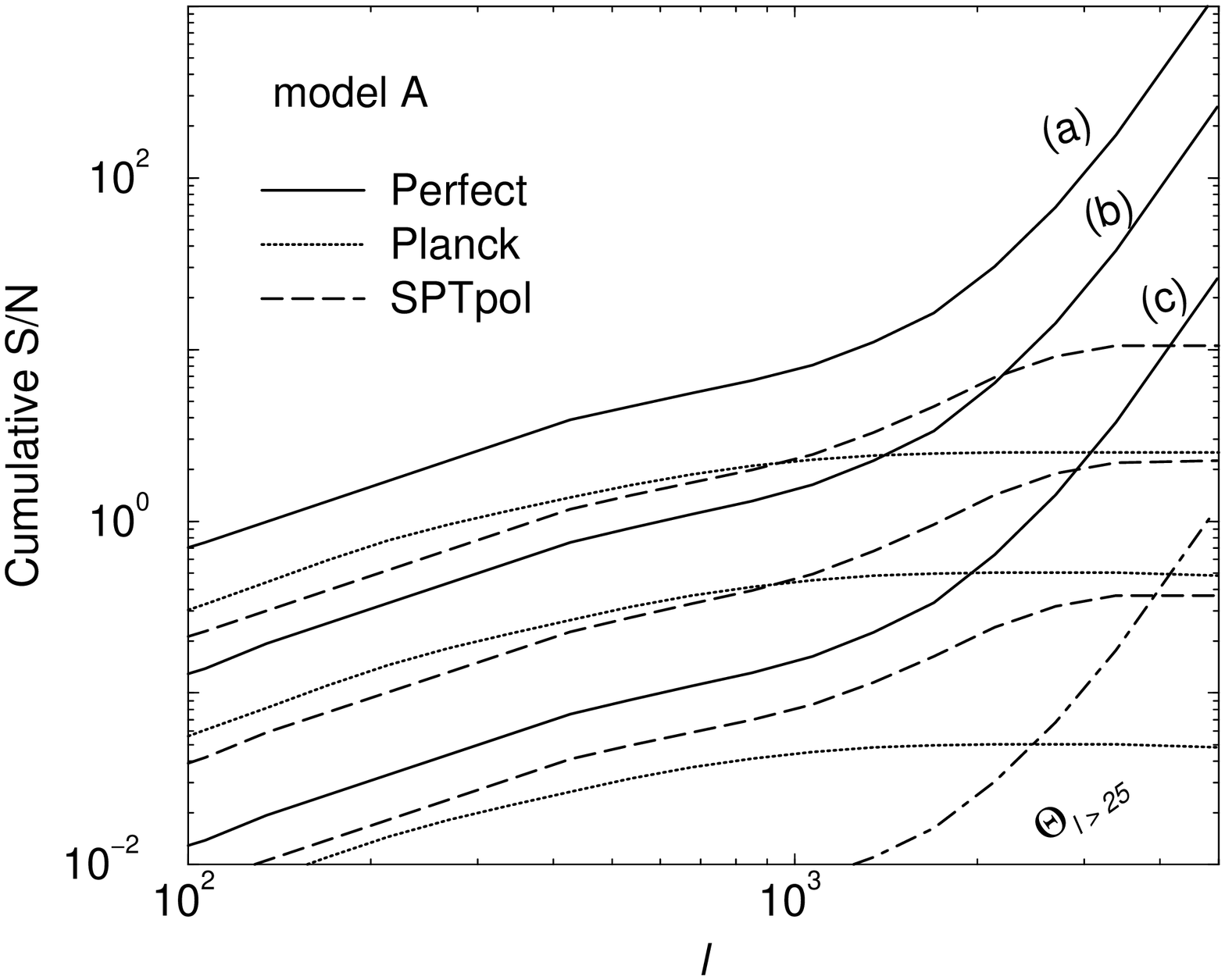,width=3.4in,angle=-90}
\psfig{file=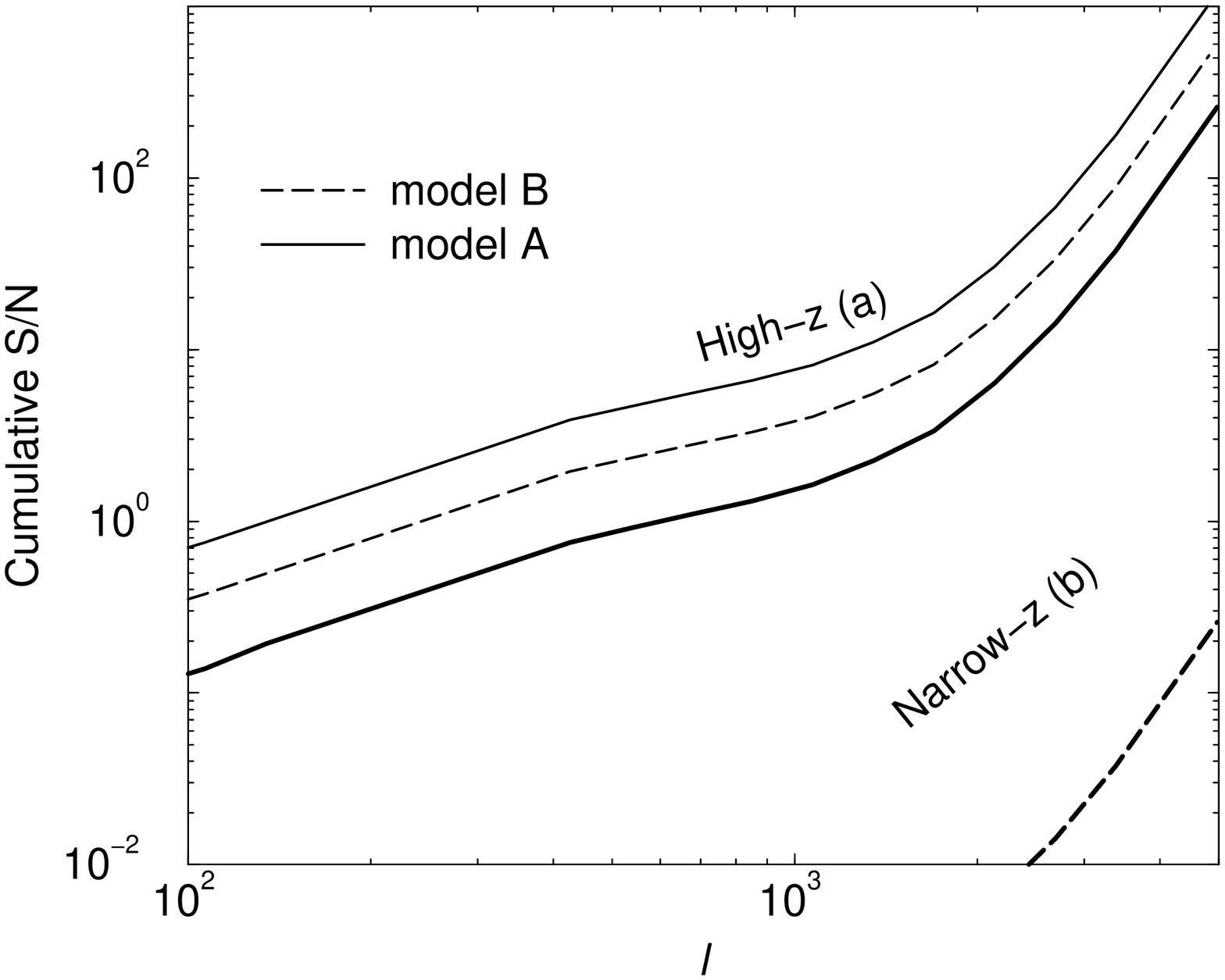,width=3.4in,angle=-90}}
\caption{The cumulative signal-to-noise for the detection of the temperature-polarization-high redshift bispectrum, as a function of
the multipole to which information is available for the high-redshift map of a tracer field of density fluctuations.In the left panel, 
we make use of model A to describe reionization history and plot
signal-to-noise ratios for Planck (all-sky; dotted lines), SPTpol (4000 sqr. degrees; dashed lines), and for a perfect experiment
limited by cosmic variance only (all-sky; solid lines). Note that for different sky-coverages, 
signal-to-noise ratios scale as $f_{\rm sky}^{0.5}$.  In each of these observational scenarios, we consider three
possibilities related to the tracer field labeled from the top to bottom as (a), (b) and (c), respectively following
Fig.~1. The bottom dot-dashed line illustrates the decrease in the signal-to-noise ratio, compared to the
to the top-most solid curve, with temperature information at multipoles greater than 25;  the correlation is dominated by
temperature  at large angular scales, and this is already available from WMAP data. 
In the right-panel, we compare the extent to which reionization histories related to models A and B can be distinguished.
Due to the lack of scattering at redshifts $\sim$ 20 in model B, when compared to model A, we find that the
cross-correlation is significantly reduced when using narrow z-band tracer fields. 
The planned upcoming studies at low radio frequencies involving the 21 cm background from neutral hydrogen
prior to around the time of reionization may provide such narrow distributions of tracer fields.}
\label{fig:sn}
\end{figure*}

As calculated in Eq.~\ref{eqn:ovbidefn}, 
the bispectrum is simply proportional to the quantity $b_{l_2l_3}$. This is an integral over the radial
distance of the redshift distribution functions related to the tracer field and electron scattering visibility, 
with their product weighted by the
cross power spectrum of fluctuations between the two. To understand how this behaves, we plot $b_{l_2l_3}$ 
in Fig.~4 as a function of $l_2$
for a given $l_3$ (top panel) and as a function of $l_3$ for certain values of $l_2$ (bottom panel). These curves generally trace the
$I_{l_2}$ integral shown in Fig.~2, but with an overall amplitude given by the strength of tracer field fluctuations and
captured with the index $l_3$. Here, we show the difference between our 
two reionization models and using the broad high-z distribution
for the tracer field. Since the reionization history related to model A is broader and expands to a higher redshift than model B,
the function $b_{l_2l_3}$ for model A, for any value of $l_3$, expands to higher values of $l_2$. 
As shown in Fig.~2, as the redshift is increased, $I(l_2)$ increases;
since this redshift traces the reionization history, 
we find that in histories where scattering happens at a higher redshift than a correspondingly
lower one, the contribution to the bispectrum increases as a function of $l_2$. 

The same difference is partly responsible for the
difference in overall amplitude of $b_{l_2 l_3}$, but as a function $l_3$ with $l_2$ fixed (lower panel of Fig.~4). Here, the overall shape of the curve as a function of
$l_3$, at a given $l_2$, is determined by the clustering strength 
and mainly the shape of the cross-power spectrum between fluctuations in the 
scattering visibility function and the tracer field. 
Incidentally, we also note that the amplitude changes between positive and negative values since the mode couping integral, $I_l$, 
oscillates between
positive and negative values. The transition from positive to negative values is simply, again, a reflection of the reionization history.
Note that the difference in curves of Fig.~4 between the two reionization models cannot simply be considered as an overall
scaling of the amplitude. The bispectra can, however, be considered as  a scaling along the $l_2$ axis for different reionization models.

\subsection{Signal-to-Noise Estimates}

In order to consider the extent to which the cross-correlations can be detected and studied to understand the reionization process, 
we estimate the signal-to-noise ratio for a detection of the bispectrum and summarize our results in Fig.~5.  
The signal-to-noise is calculated as
\begin{equation}
\left(\frac{{\rm S}}{{\rm N}}\right)^2 = \sum_{l_1 l_2 l_3}
        \frac{\left(B^{E\Theta S}_{l_1 l_2 l_3}\right)^2}{
          C_{l_1}^{\rm EE,tot}
	  C_{l_2}^{\rm CMB,tot}
          C_{l_3}^{\rm SS,tot}}\,,
\label{eqn:chisq}
\end{equation}
in the case of the E-mode related bispectrum while $E \rightarrow B$ when the B-mode bispectrum is used.
Here, $C_l^{i,tot}$ represents all contributions to the power spectrum of the $i$th field
and we write
\begin{eqnarray}
C_l^{i,tot} = C_l^i + C_l^{\rm noise} + C_l^{\rm foreg}\, ,
\label{eqn:cltot}
\end{eqnarray}
where $C_l^{\rm noise}$ is the noise contribution and $C_l^{\rm foreg}$ is the confusing foreground contribution.
We refer the reader to Ref.~\cite{Cooray:1999kg} for details on the derivation related to this signal-to-noise and
the relation between bispectrum and various other statistics at the three-point level, such as the skewness.
Instead of focusing on these statistics, which are all reduced forms of the bispectrum either in real-space or Fourier-space,
we will focus here on the bispectrum directly and consider its detection. Note that the bispectrum, as discussed in
Ref.~\cite{Cooray:1999kg}, capture all information at the three-point level and all other forms of statistics at this
level capture only a reduced amount of information with a decrease in the signal-to-noise ratio depending on the 
exact statistic used and the shape of the bispectrum. While other statistics, such as skewness in real space is more
easily measurable, we note that techniques now exist to construct the bispectrum reliably from CMB and large scale
structure maps and have been successfully applied to understand the presence of non-Gaussianities  in current data.
Thus, we do not consider the measurement of the proposed bispectrum, involving CMB temperature, polarization
and tracer-fields of high-z universe, to be any more complicated than what is already achieved \cite{Komatsu:2001wu}.

In Eq.~\ref{eqn:cltot}, when describing noise, in the case of CMB temperature maps, 
we set $C_{l_1}^{\rm CMB,tot} = C_{l_1}^{\rm CMB} + C_{l_1}^{\rm noise}$, where the
noise contribution is related to WMAP and Planck data. Since for small values of $l_1$ below few hundred,
$C_{l_1}^{\rm CMB} >> C_{l_1}^{\rm noise}$ in current data, we only need to consider cosmic variance.
For polarization maps, we take $C_{l_2}^{\rm EE,tot}= C_{l_2}^{\rm EE} + C_{l_1}^{\rm noise}+C_l^s$,
where $C_{l_1}^{\rm noise}$ is the polarization noise and $C_l^s$ is the secondary contribution to polarization
anisotropy power spectrum as shown in Fig.~3. In general, $C_{l_2}^{\rm EE} >> C_l^s$, but $C_{l_2}^{\rm EE}$ is
not significantly higher than $C_{l_1}^{\rm noise}$ for upcoming experiments such as Planck 
such that one must account for noise properly beyond just the cosmic variance. For the tracer field,
we primarily consider only the presence of cosmic variance such that 
$C_{l_3}^{\rm SS,tot}=C_{l_3}^{\rm SS}$. This is due to the fact that we do not have detailed information related to detector and/or instrumental noise
in certain high-z observations such as in the 21 cm background. 

We can roughly estimate noise 
in one case involving the planned measurement of IR fluctuations due to high-z luminous sources such as the population of
first propto-galaxies. We include this noise when considering cross-correlation studies with the high-z
broad distribution shown in Fig.~1, since such a broad redshift distribution is expected for the high-z IR bright sources.
In this case, we also include the shot-noise contribution to the tracer-field map associated with the finite density of
these sources and based on models in Ref.~\cite{Cooray:2003yf}. Since the shot-noise generally peak at scales less than an arcminute,
and the cross-correlation we are after is at angular scales of few arcminutes and more, 
we, however, find that this additional noise at small scales not to be important. 

\begin{figure}[!th]
\centerline{\psfig{file=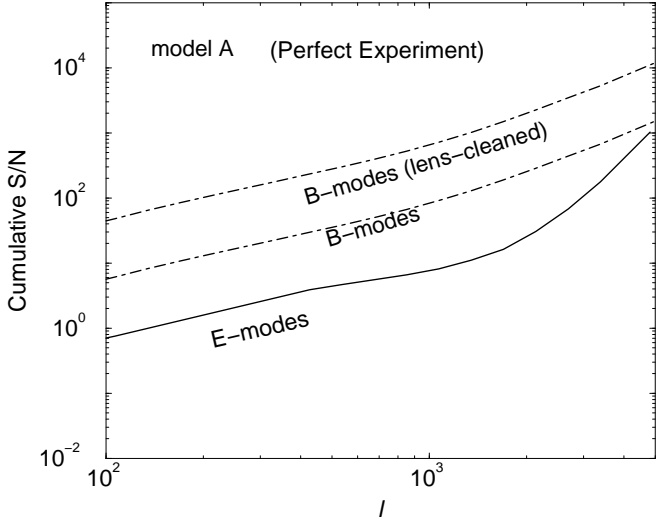,width=3.4in,angle=-90}}
\caption{The cumulative signal-to-noise for the detection of the temperature-polarization-high redshift bispectrum
using maps of the B-mode polarization and assuming a perfect experiment and a tracer field given by the (a) distribution of Fig.~1.
The top dot-dashed line is for the case where the lensing B-mode power is reduced with quadratic lensing statistics (see text for details).
For comparison, we also show the maximum E-mode signal-to-noise ratio. For other distributions considered, the signal-to-noise ratios
scale in the same manner as  shown in Fig.~5. In general, B-mode polarization maps
provide a higher signal-to-noise ratio than a map of the E-mode. Though the amplitude of the foreground reionization signal is
same in both E- and B-mode maps, due to a significant decrease in the primordial signal which contribute to the
overall variance, the cross-correlation study is better with B-modes than E-modes as the sample variance, relative to
the foreground signal, is reduced. }
\label{fig:snb}
\end{figure}

As shown in Fig.~5 (left panel), if the reionization history follows model A, 
 cumulative signal-to-noise ratios range from about a few with Planck polarization maps  to 
$\sim 10^3$ using perfect maps of the arcminute polarization field. 
These signal-to-noise ratios are substantial and are due to
the fact that the correlation between temperature and the temperature quadrupole, as well as the correlation between
visibility fluctuations and maps of the high-z universe, is high. As shown in Fig.~2, if scattering were to be restricted to
very low redshifts, $z \sim 1$, then the correlation between temperature and the temperature quadrupole is not significant
at multipoles greater than a few. On the other hand, as one moves to redshifts $\sim$ 20, this correlation is broader and spans
over few tens of multipoles. Considering this, the existence of the high signal-to-noise can be understood simply as following.
If we rewrite Eq.~\ref{eqn:bl2} for the case with a narrow redshift distribution, such as the case for high-z narrow correlation involving
21 cm fluctuations, say at a distance of $\rad_x$, we have
\begin{equation}
b_{l_1,l_2,l_3} \approx  I_{l_2}^E(\rad_x) C_{l_3}^{gS}(\rad_x)   \, ,
\end{equation}
where $C_{l_3}^{gS}(\rad_x)$ is the cross-power spectrum between fluctuations in the visibility function and that of the tracer field, $S$,
and is simply $C_{l_3}^{gS}(\rad_x) \sim r_{gS}\sqrt{C_{l_3}^{gg} C_{l_3}^{SS}}$. 
Similarly, $I_{l}^E(\rad_x)$ roughly scales as the cross-power spectrum between E-mode polarization and temperature, 
at $\rad_x$, and  we can write $I_{l_2}^E(\rad_x) \sim r_{ET} \sqrt{C_{l_2}^{EE}C_{l_2}^{TT}}$.
The signal-to-noise ratio scales as
\begin{eqnarray}
\left(\frac{{\rm S}}{{\rm N}}\right)^2 &\approx& \sum_{l_1 l_2 l_3} f(l_1,l_2,l_3)
\frac{r_{ET}^2 r_{gS}^2 C_{l_2}^{EE}  C_{l_3}^{gg}}{
  C_{l_1}^{\rm EE,tot}} \,, 
\label{eqn:sn}
\end{eqnarray}
where $f(l_1,l_2,l_3) \sim l_1^2 l_2^2 l_3^2 \wj^2$. The summation over $l_3$, $\sum_{l_3} l_3^2 C_{l_3}^{gg}$ is simply the
variance related to visibility fluctuations and increases as the maximum value of $l$ over which $l_3$ is summed.
The ratio of $C_{l_2}^{EE}/(C_{l_1}^{\rm EE,tot})$
behaves on the signal-to-noise associated with polarization measurements, especially
when $l_1 \sim l_2$. Thus, roughly, signal-to-noise in the correlation bispectrum, 
for each $l_1$ and $l_2$ mode, scales as $f_{\rm sky}^{1/2} 
r_{ET}^2 r_{gS}^2 \sigma^2_{gg} ({\rm S}/{\rm N})^2_{\rm pol}$ and is determined mostly by the resolution of the
polarization map and the extent to which fluctuations in the visibility function is correlated with that of the
tracer field. Since $r_{ET}^2 \sim 1$ and $\sigma_{gg}$ is $\sim$ 0.6 percent (when $l_3 \sim 1000$)\footnote{This can be understood based on
the fact that the secondary polarization due to visibility fluctuations shown in Fig.~3 is simply $C_l \sim 3/100 C_l^{gg} Q_{\rm rms}^2$ and
since $Q_{\rm rms} \sim 25$ $\mu$K,  $l^2 /2 \pi C_l^{gg}$ is of order $4 \times 10^{-5}$ or $\sigma_{gg} \sim 0.6$\%},
if $r_{gS}^2 \sim 1$, one can achieve, in principle, signal-to-noise values of order $\sim$ few hundred with all-sky maps
of polarization and adequate resolution in polarization down to multipoles of 1000.
This, of course, assumes no noise maps in both polarization and in the tracer field and just that the noise is limited by the cosmic variance.
In Fig.~5, we include instrumental and observational noise which lead to a decrease in the signal-to-noise ratio, for example, with Planck by
several orders of magnitude.

In Fig.~5 (right panel), we illustrate the difference between expected signal-to-noise ratios for reionization histories
involving models A and B. Here, we consider two tracer-fields involving the high-z broad distribution and the
high-z narrow distribution. Fig.~1 shows the overlap between these distributions and the 
 visibility function related to the two reionization models. This overlap, or its non-existent,
helps understand differences in the bispectrum signal-to-noise ratios.
In particular, we note the sharp reduction in the signal-to-noise ratio between the two models 
when using the narrow redshift distribution for tracer-field at
high redshifts from model A to model B. This is due to the fact that most, if not all, rescattering is concentrated at
redshifts below 15 in the case of model B while the tracer field is a probe of structure at redshifts of order 20. 

The situation is also the same when one uses a tracer-field with a broad redshift-distribution,
 but limited to redshifts greater than 15.
While this allows a mechanism to distinguish the extent to which two broadly different reionization models can be
distinguished form one another, though they both give the same optical depth and essentially produce the same polarization signature,
one can do more than this. Under the assumption that the reionization history follows an extended period, one can use
tracer-fields with narrow band distributions to study the strength of the correlation as a function of redshift. This in
return can be converted as a measurement of the visibility function as a function of redshift. While we have not considered
such a possibility, due to the lack of information related to how well narrow band distributions can be defined at
redshifts of order 15, we emphasize that 21 cm observations of the neutral Hydrogen content at high redshift may be the ideal
way to approach such a reconstruction; Due to the line emission, redshift ranges can be a priori defined at observational wavelengths
and such observations, when combined with CMB, may provide an interesting possibility for detailed studies
related to reionization.

Instead of E-modes, as mentioned, we can also consider the B-mode for a cross-correlation study (Fig.~6). While there is
no large angular scale signal related to rescattering in the B-mode map, associated with density fluctuations, at tens of arcminute
scales and below, a B-mode signal exists due to modulations of the visibility function by density inhomogeneities in the ionized electron
distribution. For the B-mode bispectrum, to describe noise, we take $C_{l_2}^{\rm BB,tot}= C_{l_2}^{\rm BB} + C_{l_1}^{\rm noise}+C_l^s$
and set $C_{l_2}^{\rm BB}$ as that due to gravitational lensing conversion power from E- to B-modes, as shown in Fig.~3,
and ignore the presence of primordial gravity waves or tensor modes. This is  a safe assumption given that
such primordial contributions are expected to peak at angular scales of few degrees, while the proposed study involve
tens of arcminute scale polarization. 

Since the ratio of $C_{l_2}^{\rm sec}/(C_{l_1}^{\rm BB})$ is higher than that related to E-modes, one generally obtains a
higher signal-to-noise ratio with an arcminute scale B-mode map when compared  to a map of E-modes. The same
is expected since the primordial fluctuations are smaller in B-modes than in E-modes, while the secondary contribution
remains the same, the confusion generated by primary fluctuations for the cross-correlation is reduced in the case
of B-modes relative to the same in E-modes. This can also be stated in terms of the reduced sample variance.
There is another advantage with the use of B-mode
information. Since the main confusion at arcminute scales is related to that associated with gravitational lensing confusion,
one can improve the signal-to-noise ratios for the bispectrum measurement significantly, when the B-mode map is a priori
cleaned with quadratic statistics or improved likelihood techniques \cite{Okamoto:2003zw}. These techniques lower the B-mode
power related to lensing by more than an order of magnitude and reduces the sample variance contribution to the noise
associated with the bispectrum by a similar number.

\begin{figure}[!h]
\centerline{\psfig{file=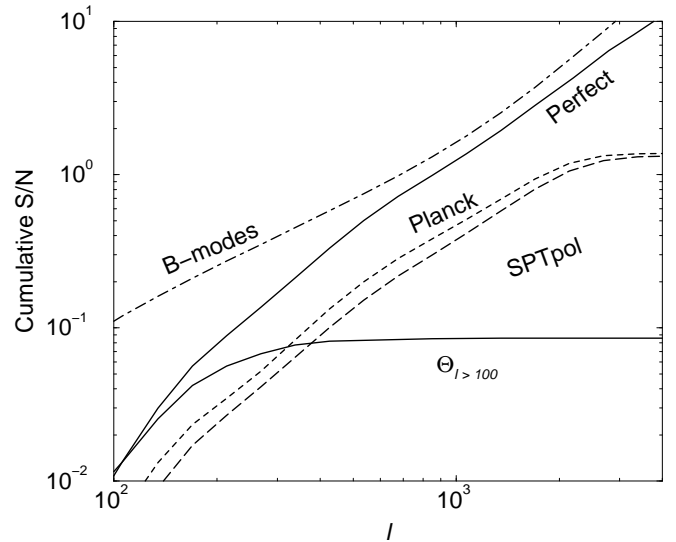,width=3.4in,angle=-90}}
\caption{The cumulative signal-to-noise for the detection of the temperature-polarization-high redshift bispectrum, 
related to the gravitational lensing effect on CMB temperature and polarization anisotropies. The non-Gaussian signal is
generated due to the fact that the high-redshift tracer field may correlate with the lensing deflection potential
that is modifying the spatial distribution of CMB fluctuations. We show the signal-to-noise ratio for the tracer field
involving the high-z broad distribution and for a perfect, Planck  and SPTpol like ground-based experiment.
The top curves are for the case where all information from the temperature map is included while the bottom curve, for a perfect
polarization measurement, is for the case where we restrict temperature information to a multipole of 100 only. While the reionization related
bispectrum is not affected by such a cutoff, due to the fact that relevant scales related to the correlation comes from
large angular scale temperature anisotropy, in the case of lensing, this results in a significant reduction of this signal-to-noise.
The dot-dashed line shows the signal-to-noise ratio associated with a bispectrum measurement related to lensing
using the B-mode map, in addition to temperature and tracer-field information. 
Note the difference in scale of the ordinate between Figs.~5 and 6 and this plot.}
\label{fig:lenssn}
\end{figure}

\subsection{Contaminant Correlations}

Using Figs.~5 and 6, we have argued that there is sufficient signal-to-noise for the observational measurement of the proposed correlation in the form of
a bispectrum. It is also useful to consider if this bispectrum can be confused with other sources of non-Gaussianities in the CMB temperature
and polarization anisotropies when correlated in the same manner with a map of the high redshift universe. The only possibility
is related to the weak lensing effect on CMB anisotropies \cite{Hu:ee}. Essentially, weak lensing deflections of CMB photons lead to a
correction to CMB anisotropy that depends on the temperature gradient and the deflection angle. While the temperature gradient
correlates with the E-mode polarization map, the deflection angle can correlate with a map of fluctuations in the high-z universe, and generate a
non-Gaussian signal related to the bispectrum.  

We write the bispectrum related to this cross-correlation as
\begin{eqnarray}
&&B^{E\Theta S}_{l_1 l_2 l_3}  
=\frac{1}{2}\sqrt{\frac{(2l_1 +1)(2 l_2+1)(2l_3+1)}{4 \pi}} \nonumber \\
&&\Big\{\left(
\begin{array}{ccc}
l_1 & l_2 & l_3 \\
2 & 0  &  -2
\end{array}
\right) F(l_1,l_2,l_3) C_{l_2}^{\Theta E} C_{l_3}^{\phi S} \nonumber \\
&+& \left(
\begin{array}{ccc}
l_1 & l_2 & l_3 \\
0 & 0  &  0
\end{array}
\right) F(l_2,l_3,l_1)  C_{l_1}^{\Theta E} C_{l_3}^{\phi S} \Big\}\, , \nonumber \\
\label{eqn:lensbidefn}
\end{eqnarray}
where $F(l_1,l_2,l_3) = \left[l_2(l_2+1)+l_3(l_3+1)-l_1(l_1+1)\right]$
$C_l^{\phi S}$ is the cross-power spectrum between lensing potentials and the tracer field
and is given by
\begin{equation}
C_l^{\phi S} = \int dr \frac{W^S(\rad)}{\da^2} W^\phi\left(k=\frac{l}{\da},\rad\right) P_{\delta S}\left(k=\frac{l}{\da}\right) \, ,
\end{equation}
where 
\begin{equation}
W^\phi(k,\rad) = -3 \Omega_m \left(\frac{H_0}{k}\right)^2 F(\rad) \frac{\da(\rad_0-\rad)}{\da(\rad)\da(\rad_0)} \, .
\end{equation}

Similarly, lensing induces a bispectrum related to B-modes combined with temperature information and a tracer-field.
We can write this bispectrum as
\begin{eqnarray}
&&B^{B\Theta S}_{l_1 l_2 l_3}  
=\frac{i}{2}\sqrt{\frac{(2l_1 +1)(2 l_2+1)(2l_3+1)}{4 \pi}} \nonumber \\
&&\left(
\begin{array}{ccc}
l_1 & l_2 & l_3 \\
2 & 0  &  -2
\end{array}
\right) F(l_1,l_2,l_3) C_{l_2}^{\Theta E} C_{l_3}^{\phi S} \, . \nonumber \\
\label{eqn:lensbiBdefn}
\end{eqnarray}
In general, for high-z distributions of the tracer-field, $W^S(\rad)$ and $W^\phi$ do not overlap significantly. This results in
a reduction in the cross-correlation between the two. Moreover, since the reionized polarization anisotropy related bispectrum can
be constructed with $\Theta_l < 100$ or so without a loss of signal-to-noise,  the importance of the lensing bispectrum can be
significantly reduced since limiting the lensing bispectrum to $l_2$ values less than 100, leads to
a  substantial reduction in the signal-to-noise associated with it. Even with a perfect experiment,
and $l_2$ multipole taking same values as $l_3$ out to 1000 or more,
the signal-to-noise ratios are not greater than few tens \cite{Hu:ee} (see, Fig.~7). With $l_2$ reduced to a smaller value below 100,
this signal-to-noise ratio decreases substantially below a few, at most. Given that the polarization fluctuations related to
bispectrum has substantially higher signal-to-noise ratios, it is unlikely that the lensing generated non-Gaussianities are
a major source of concern for the proposed study related to reionization.

\section{Summary}
\label{sec:summary}

The free-electron population during the reionized epoch rescatters CMB temperature quadrupole and generates
a now well-known polarization signal at large angular scales. While this contribution has been detected in
the temperature-polarization cross power spectrum from WMAP data, due to the large cosmic variance 
associated with anisotropy measurements at relevant scales,
only limited information related to reionization, such as the optical depth to electron scattering, can be extracted.
The inhomogeneities in the free-electron population lead to an additional secondary polarization anisotropy contribution 
at arcminute scales. 

While the fluctuation amplitude, relative to dominant primordial fluctuations, is small, 
we suggest that a cross-correlation between arcminute scale CMB polarization data and a tracer field of the high redshift universe, 
such as through fluctuations captured by the 21 cm neutral Hydrogen background or those in the
infrared background related to first proto-galaxies, may allow one to study additional details related to reionization.
This includes a possibility to recover the visibility function 
related to electron scattering. For this purpose, we discuss an optimized higher-order
correlation measurement, in the form of a three-point function, involving
information from large angular scale CMB temperature anisotropies in addition to 
arcminute scale polarization data.  The proposed bispectrum can be measured with a substantial signal-to-noise ratio
with lens-cleaned B-mode maps where the confusion related to primordial anisotropies and the arcminute scale reionization-related
polarization is the least; this can be compared to other sources of
non-Gaussianity in CMB data, such as due to gravitational lensing, which with the same combination of temperature, polarization,
and the tracer field is at the level of at most unity in terms of the signal-to-noise ratio. 

Given that the arcminute
scale polarization fluctuations are correlated with fluctuations at high redshift at same angular scales, for a reliable
measurement of the three point correlation function suggested here, one  
does not require all-sky maps of CMB polarization or that of the tracer field. This is helpful since one can obtain high signal-to-noise
maps of the polarization from ground-based experiments, 
and in certain cases fluctuation data on certain tracers, by concentrating on limited sky-area instead of the
whole sky. The required information from temperature anisotropies is at large angular scales and this information is already available
to the limit allowed by the cosmic variance with WMAP data and is expected to improve with Planck in terms of the foreground confusion
and separation. A study such as the one proposed may allow one to establish the epoch when CMB polarization related to reionization is generated and to address if the universe was reionized once or twice.

\acknowledgments
This work was supported in part by DoE DE-FG03-92-ER40701 and 
a senior research fellowship from the Sherman Fairchild foundation.

\end{document}